\documentclass[aps,superscriptaddress,fleqn,showpacs,pre]{revtex4}
\usepackage{amsmath,amssymb}
\usepackage{mathrsfs}
\usepackage{graphicx}
\graphicspath{{.},{../revise1_20140223/},{../submit20140117/}}
\usepackage{bm}
\usepackage{color}

\newcommand{\ii}{\mathrm{i}}
\newcommand{\re}[1]{#1_\mathrm{R}}
\newcommand{\im}[1]{#1_\mathrm{I}}
\newcommand{\e}{\epsilon}

\begin{document}



\title{Clustering in Globally Coupled Oscillators Near a Hopf Bifurcation:
Theory and Experiments }

\author{Hiroshi Kori}
\email[corresponding author: ]{kori.hiroshi@ocha.ac.jp}
\affiliation{Department of Information Sciences, Ochanomizu Univeristy, Tokyo 112-8610, Japan}
\affiliation{CREST, Japan Science and Technology Agency, Kawaguchi
332-0012, Japan}
\author{Yoshiki Kuramoto}
\affiliation{International Institute for Advanced Studies, Kyoto 619-0225, Japan}
\author{Swati Jain}
\affiliation{Department of Chemical Engineering, University of Virginia,
CharlottesVille, Virginia 22904, USA}
\author{Istv\'{a}n Z. Kiss}
\affiliation{Department of Chemistry, Saint Louis University, St. Louis,
Missouri 63103, USA}
\author{John Hudson}
\affiliation{Department of Chemical Engineering, University of Virginia,
CharlottesVille, Virginia 22904, USA}


\date{\today}

\begin{abstract}
A theoretical analysis is presented to show the
general occurrence of phase clusters in weakly, globally coupled oscillators
close to a Hopf bifurcation. 
Through a reductive perturbation method, we derive the amplitude
equation with a higher order correction term
valid near a Hopf bifurcation point.
This amplitude equation allows us to calculate analytically the phase
coupling function from given limit-cycle oscillator models.
Moreover, using the phase coupling function, the stability of phase
clusters can be analyzed. 
We demonstrate our theory with the Brusselator model.
Experiments are carried out to confirm the presence of phase clusters close to
Hopf bifurcations with electrochemical oscillators. 
\end{abstract}
\maketitle

\section{Introduction}
The dynamics of weakly interacting oscillating units can be described
with phase models \cite{kuramoto84}. For example, for $N$ identical oscillators with
global (i.e., mean field) coupling, we have
\begin{equation}
 \frac{d\phi_{i}}{dt}=\omega+\frac{\kappa}{N} \sum_{j=1}^N\Gamma(\phi_{i}-\phi_{j}),
  \label{pm}
\end{equation}
where $\phi_{i}$ $(i=1,\ldots,N)$ and $\omega$ are the phase
and the frequency of oscillator $i$, respectively, 
$\kappa$ is the coupling strength, 
and $\Gamma$ is
the phase coupling function.
Phase models have been formulated from 
ordinary differential equations describing, e.g., chemical reactions \cite{kuramoto84}.
Recently, they have been formulated from direct experiments as well
\cite{kiss05, kiss07, miyazaki06}.
A prominent feature
of such phase models is the presence of higher harmonics in the phase coupling
function. As a result of higher harmonics, complex dynamics including
chaos and multi-phase clusters can be
observed \cite{okuda93, hansel93, kori01, kori03,bick11}.
Here, we refer to phase clusters as clustering behavior purely
attributed to phase dynamics. Clustering dynamics that can not be described by
phase models are refereed to amplitude clusters \cite{nakagawa93}. 
Phase clusters have been experimentally observed in a wide range
of systems including electrochemical systems, light sensitive BZ reaction,
carbon monoxide oxidation on platinum \cite{yang00, vanag00, kim01, wang02, kiss05,kiss07,kori08, varela05, taylor11}. The number of clusters and
their appearance with positive or negative coupling/feedback is puzzling.
Multi-phase clusters are usually explained by the presence of higher harmonics
in the coupling functions in the phase model \cite{okuda93}; however, the
mechanism through which higher harmonics can develop is unclear. 

In this paper, we give a theoretical explanation for clusters close
to oscillations that develop through Hopf bifurcations. With analytical
derivation, we show how the higher harmonics occur in the phase coupling
function through two-step reductions: an amplitude equation is
derived from coupled limit-cycle oscillators
through a reductive perturbation method,
and then, the phase coupling function is derived from
the amplitude equation through the phase reduction method.
Experiments are carried out to confirm the presence of multi-phase clusters close to
Hopf bifurcations with electrochemical oscillators. 

\section{theory} \label{sec:theory}
\subsection{Overview} \label{sec:overview}
The Stuart-Landau (SL) oscillator, which is a local element of the complex
Ginzburg-Landau model\cite{aranson02,garcia12}, 
is considered as a general skeleton
model for the description of oscillators close to Hopf bifurcations.  It
can be derived as a lowest-order amplitude equation through a reductive perturbation method
\cite{kuramoto84}, as summarized in Sec.~\ref{sec:lowest}.
However, a population of SL oscillators coupled
globally and linearly can not describe multi-phase clusters because its
corresponding phase coupling function $\Gamma$ in Eq.~\eqref{pm}  does not contain second or higher harmonics; i.e.,
$\Gamma(\Delta\phi) \propto \sin(\Delta\phi+\theta)+a_0$, where $a_0$ and $\theta$
are constant \cite{kuramoto84,nakagawa93}. 
As shown in Table \ref{table:theory}a, 
the stability analysis of such a phase model predicts that the only
stable behavior is the one cluster state (in-phase synchrony).
\begin{table}[tb]  
\begin{center}
\begin{tabular}{|c|c|c|c|}
\hline
 (a)  & 1 cluster & Desync & Multi-clusters \\
\hline
 $\kappa > 0$ & {stable} & unstable &  unstable \\\cline{1-3}
 $\kappa < 0$ & unstable & neutral & or neutral \\
\hline
\end{tabular}
\hspace{5mm}
\begin{tabular}{|c|c|c|c|}
\hline
 (b)  & 1 cluster & Desync & Multi-clusters \\
\hline
 $\kappa > 0$ & {stable} & unstable &  unstable, \\\cline{1-3}
 $\kappa < 0$ & unstable & unstable or  neutral& neutral, or {stable} \\
\hline
\end{tabular}
\caption{General properties of phase clusters in theoretical models
close to a Hopf bifurcation with weak, global coupling. (a) Properties
in the lowest order amplitude equation [Eq.~\eqref{gl}]. (b) Properties in
the higher order amplitude equation [Eq.~\eqref{gl2}]. 
``Desync'' refers to the desynchronized state,
which is the state of uniform phase distribution. This state is also
called the splay state.
Here we assumed that the one cluster state is asymptotically stable
for $\kappa>0$. When one cluster state is stable for $\kappa<0$,
the stability of the desynchronized state is exchanged
between $\kappa>0$ and $\kappa<0$.
}
\label{table:theory}
\end{center}
\end{table}

However, it should be noticed that infinitesimally small perturbations
given to the Stuart-Landau system may alter a neutral state to a stable or unstable
one; i.e., the Stuart-Landau system is structurally unstable.
Indeed, we will show that when we take into account higher order
correction to
the amplitude equation, multi-cluster states may become
asymptotically stable.
The number of clusters and whether they occur with negative or positive
coupling depends on the types of nonlinearity in the ordinary differential
equations. The stability of the cluster state is weak, nonetheless it is expected
that they can be observed in globally coupled oscillators even
very close to a Hopf bifurcation point.


\subsection{Lowest order amplitude equation} \label{sec:lowest}
Consider a network of identical limit-cycle
oscillators with linear coupling
\begin{equation}
\dot{\bm x}_i=\bm f (\bm x_i ; \mu) + \mu \kappa 
 \sum_{j=1}^N  A_{ij} \hat D (\bm x_j-\bm x_i),\label{full_model}
\end{equation}
where $\bm x_i\in \mathbb{R}^M$ is the state variable of the $i$-th oscillator
($i=1,\ldots,N$), $\hat D$ is a $M \times M$ matrix,
and $\mu \kappa$ is the coupling strength. 
We assume $\bm f (\bm x = \bm 0; \mu) = \bm 0$ without loss of generality.
We also assume that, in the absence of coupling ($\kappa=0$),
the trivial solution $\bm x_i=\bm 0$ is stable for $\mu<0$ and
undergoes a supercritical Hopf bifurcation at $\mu=0$, such that
each unit becomes a limit-cycle oscillator with the amplitude $|\bm x_i|=O(\sqrt{\mu})$ for $\mu>0$. Hereafter, we consider only $\mu \ge 0$
and put $\mu = \epsilon^2$ for convenience.


To derive the amplitude equation for Eq.~\eqref{full_model} near a Hopf bifurcation,
it is sufficient to focus on the subsystem in which an oscillator
is coupled to another. Let $\bm x$ and $\bm x'$ be the state vectors of
the oscillators. The dynamical equation for $\bm x$ is given as
\begin{equation}
\dot{\bm x}=\bm f (\bm x ; \epsilon^2) + \epsilon^2 \kappa \hat D \bm x'.\label{full_model2}
\end{equation}
We expand $\bm f(\bm x; \e^2)$ around $\bm x=\bm 0$
as
\begin{equation}
 \bm f(\bm x;\epsilon^2)=\bm{n}_1(\bm{x}; \epsilon^2) 
  +\bm{n}_2(\bm{x},\bm{x}; \epsilon^2)+\bm{n}_3(\bm{x},\bm{x},\bm{x}; \epsilon^2)+O(|\bm x|^4),
\end{equation}
where $\bm n_k$ ($k=1,2,3$) is the $k$-th order term
in the expansion
(the precise definitions of $\bm n_2$ and $\bm n_3$ are given in Appendix~\ref{sec:appA}).
We further expand $\bm n_k$ with respect to $\epsilon^2$ to obtain 
\begin{equation}
 \bm f(\bm x;\epsilon^2)=\hat L_0 \bm{x}+\e^2 \hat L_1
  \bm{x}+\bm{n}_2(\bm{x},\bm{x})+\bm{n}_3(\bm{x},\bm{x},\bm{x})+O(|\bm
  x|^4),
\label{expansion}
\end{equation}
where $\bm n_k(\cdot)$ ($k=1,2$) denotes $\bm n_k(\cdot;\epsilon^2=0)$. 
Note that $O(\e^2)$ terms in $\bm{n}_2$ and $\bm{n}_3$ are irrelevant
to the calculations below and thus omitted in Eq.~\eqref{expansion}.
Because of the assumption of Hopf bifurcation, $\hat L_0$ has a pair of purely imaginary eigenvalues $\pm\ii \omega_0$. 
The right and left eigenvectors of $\hat L_0$ corresponding to the
eigenvalue $\ii \omega_0$ are denoted by $\bm u$ (column vector) and
$\bm v$ (raw vector), respectively; i.e., $\hat L_0 \bm u = \ii \omega_0
\bm u$ and $\bm v \hat L_0 = \ii \omega_0 \bm v$.
They are normalized as $\bm v \bm u=1$.
The solution to the linearized unperturbed system, $\dot{\bm x} = L_0 \bm{x}$, is
given by
\begin{equation}
\bm{x}_0(t)= w e^{\ii\theta(t)} \bm{u}+\bar w e^{-\ii\theta(t)} \bar{\bm u},
\end{equation}
where $w$ is an arbitrary complex number, which we refer to as the complex amplitude;
$\bar{\bm u}$ and $\bar w$ denote the complex conjugate of $\bm u$ and
$w$, respectively; and $\theta(t)=\omega_0 t$.

In 
Eq.~\eqref{full_model}, $\bm x(t)$
generally deviates from $\bm x_0(t)$.
By interpreting $w$ as a time-dependent variable $w(t)$,
it is possible to describe the
time-asymptotic behavior of $\bm x(t)$ in the following form
\begin{align}
 \bm x &= \bm x_0(w, \bar w, \theta) + \bm \rho(w, \bar w, w',\bar w',\theta), \label{reduction_x}\\
 \dot w &= g(w,\bar w) + \e^2 \kappa h(w, \bar w, w',\bar w'), \label{reduction_w}
\end{align}
where $\bm \rho, g$ and $h$ are the functions to be determined
perturbatively.
Note that $g$ and $h$ are free from $\theta(t)$.
Equation
\eqref{reduction_w} is called the amplitude equation.
In Ref. \onlinecite{kuramoto84}, the amplitude equation to the lowest order
is derived as
\begin{equation}
 \dot w = \e^2 \alpha w - \beta |w|^2 w + \e^2 \kappa \gamma w',
  \label{gl}
\end{equation}
where $\alpha,\beta$ and $\gamma$ are the complex constants with the
following expressions:
\begin{align}
 \alpha &= \bm v \hat L_1 \bm u, \label{alpha}\\
 \beta  &=  - 3\bm v \bm n_3(\bm u,\bm u,\bar{\bm u})
  + 4 \bm v \bm n_2(\bm u,\hat L_0^{-1}\bm n_2(\bm u,\bar{\bm u}) )
  + 2 \bm v \bm n_2(\bar{\bm u},(\hat L_0 - 2 \ii \omega_0)^{-1} \bm n_2(\bm u,\bm u)), \label{beta} \\
 \gamma &= \bm v \hat D \bm u. \label{gamma}
\end{align}

\subsection{Phase reduction for the lowest order amplitude equation}
Following the method in Ref. \onlinecite{kuramoto84}, we may further reduce the amplitude equation to a phase model.
For $\kappa=0$, Eq.~(\ref{gl}) has the stable limit-cycle solution given by
\begin{equation}
 w_0(t) = r e^{\ii \phi(t)}
  \label{w_0}
\end{equation}
where $r=\e \sqrt{\re\alpha/\re\beta}$ (the
subscripts R and I denote the real and imaginary parts, respectively),
$\phi(t) = \omega t + \phi_0$, $\omega = \e^2 \re \alpha (c_0-c_2), c_0=\im
\alpha/\re \alpha$, $c_2= \im \beta/\re \beta$, and $\phi_0$ is an arbitrary
initial phase. For sufficiently small $\kappa$,  
the trajectory of $w(t)$ deviates only slightly from that of the unperturbed limit-cycle. 
In this case, $w(t)$ is well approximated by
$r e^{\ii \phi(t)}$ with the phase $\phi(t)$ obeying the following phase model:
\begin{equation}
 \dot \phi = \omega + \e^2 \kappa \Gamma(\phi-\phi'). \label{pm2}
\end{equation}
The phase coupling function $\Gamma$ is obtained through
\begin{equation}
 \Gamma(\phi-\phi') = \langle z(\phi) \cdot h(w_0,w'_0)\rangle,
  \label{Gamma}
\end{equation}
where $z(\phi)$ is the phase sensitivity function, $a \cdot b = 
(\bar a b + a \bar b)/2$ denotes the inner product in a complex form,
and $\langle f(\phi,\phi') \rangle = \frac{1}{2\pi} \int_0^{2\pi} f(\phi+\mu,
\phi'+\mu) d\mu$ denotes averaging.
The phase sensitivity function is determined by the nature of
limit-cycle oscillation. In the case of Eq.~\eqref{gl}, 
we have \cite{kuramoto84}
\begin{equation}
 z(\phi) = \frac{-c_2 + \ii}{r} e^{\ii \phi}.
  \label{z_gl}
\end{equation}

For convenience, we expand the phase coupling function $\Gamma$ as
\begin{equation}
 \Gamma(\psi)=a_0+\sum_{\ell=1}^\infty( a_\ell \cos \ell\psi + b_\ell \sin \ell\psi).
\end{equation}
Substituting Eq.~\eqref{z_gl} and $h(w_0,w'_0)=\gamma w'_0 =  \gamma r e^{\ii \phi'}$ to Eq.~\eqref{Gamma}, we find
\begin{align}
 a_1 &= \re \gamma(c_1-c_2),\label{a_1}\\
 b_1 &= -\re \gamma (1+c_1 c_2), \label{b_1}
\end{align}
where $c_1 \equiv \im \gamma/ \re \gamma$.
Importantly, all other coefficients vanish.

Note that, if the coupling term in Eq.~\eqref{full_model2} is diffusive
[i.e., $\epsilon^2 \kappa \hat D (\bm x'-\bm x)$ instead of $\epsilon^2 \kappa \hat D \bm x'$],
we have $h(w_0,w'_0)=\gamma (w'_0-w_0)$.
In this case, we obtain $a_0 = -a_1$ in
addition to Eqs.~\eqref{a_1} and \eqref{b_1} .

We have seen that, for the lowest order amplitude equation given by
Eq.~\eqref{gl}, the corresponding phase coupling function does not contain
the second and higher harmonics.
Therefore, as briefly mentioned in Sec.~\ref{sec:overview}, 
this amplitude equation does not admit multi-phase clusters.
Note that strong coupling may result in amplitude clusters \cite{nakagawa93}.

\subsection{Higher order correction} \label{sec:higher_order}
Now, we derive higher order correction terms to the amplitude equation and 
the corresponding phase coupling function $\Gamma$.
We here consider only weak coupling; i.e. the corrections of $O(\kappa^2)$
are neglected. We also consider only linear coupling, as given in
Eq.~\eqref{full_model}. Our result would change if we consider
nonlinear coupling.

At first, we discuss higher order terms in $g(w)$ in Eq.~\eqref{reduction_w}.
We know that $g(w)$ consists only of $|w|^n w$ ($n=0,1,2,\ldots$), called
the resonant terms \cite{guckenheimer}.
The dynamical equation $\dot w = g(w)$ is invariant under the
transformation $w \to w e^{\ii \phi}$; i.e., the system has the
rotational symmetry. This implies that
$w_0(\phi)$ and $z(\phi)$ have the following forms:
$w_0(\phi) = w_0(0) e^{\ii \phi}$ and $z(\phi) = z(0) e^{\ii \phi}$. 
Then, obviously, $\Gamma(\phi-\phi') = \langle z(\phi) \cdot \gamma w'_0(\phi')\rangle$
contains only the first harmonics.
Note that, however, the term $|w|^n w$ provides the corrections 
of $O(\e^3)$ and $O(\e^1)$ to $w_0(\phi)$ and $z(\phi)$,
respectively. 
These corrections give rise to the corrections of $O(\e^2)$ in $a_1$ and $b_1$.


Therefore, for $\Gamma$ to possess higher harmonics,
we need to consider higher order correction to $h(w, w')$.
In general, $h(w, w')$ is described as a polynomial of $w, \bar w, w', \bar w'$. 
Let us suppose that 
$h(w, w') = w^{\ell_1} \bar w^{\ell_2}  w'^{\ell_3} \bar
w'^{\ell_4}$ with integers 
$\ell_1, \ell_2,\ell_3,\ell_4 \ge 0$. Because $w_0 = O(\e)$ and $z =
O(\e^{-1})$ [see Eqs.~\eqref{w_0} and \eqref{z_gl}], we have $z \cdot h = O(\e^{\ell_1+ \ell_2+\ell_3+\ell_4-1})$.
We also have $z \cdot h(w, w') \propto e^{\ii
(\ell_1-\ell_2-1)\phi} e^{\ii (\ell_3-\ell_4)\phi'}$.
This term contributes to $a_\ell$ and $b_\ell$ ($\ell > 0$)
when this term is a function of only $\pm \ell(\phi - \phi')$; i.e.,
$\ell_1-\ell_2-1= \pm \ell$ and $\ell_3-\ell_4 = \mp \ell$.
Obviously, $\bar w^{\ell-1} w'^\ell = O(\e^{2\ell-1})$ gives a leading contribution to $a_\ell$ and
$b_\ell$. We thus find
\begin{equation}
 a_\ell, b_\ell = O(\e^{2(\ell-1)}).
  \label{order}
\end{equation}
Other terms in $h(w,w')$ together with higher order terms in
$g(w)$ provide minor corrections to the coefficients $a_\ell$ and
$b_\ell$.

Let us focus on the resonant terms of $O(\e^3)$
in $h(w,w')$.
There are five resonant terms:
$\bar w w'^2$, $|w|^2 w'$, $w^2
\bar w'$, $|w'|^2 w'$, and $w |w'|^2$ . As already discussed, the
first term yields the second harmonic of $O(\e^2)$, thus providing
leading terms of $O(\e^2)$ to $a_2$ and $b_2$. The other resonant terms yield minor
corrections. Namely, the next three resonant terms yields
the first harmonic of $O(\e^2)$, and the
final resonant term yields the zeroth harmonic of $O(\e^2)$.
Therefore, only the term $\bar w w'^2$ gives a major
effect on phase dynamics.
Thus, as a second order amplitude equation for weakly
coupled oscillators near a supercritical Hopf bifurcation point, 
it is appropriate to consider the following equation:
\begin{equation}
 \dot w = \e^2 \alpha w - \beta |w|^2 w + \e^2 \kappa (\gamma w' +
  \delta \bar w w'^2),
  \label{gl2}
\end{equation}
where $\delta$ is a complex constant.
One of the main results in the present paper is,
as shown in Appendix \ref{sec:appA}, 
the derivation of the expression for $\delta$, which has the following
concise form: 
\begin{align}
 \delta &= 2 {\bm v} \bm n_2(\bar{\bm u},
             (\hat L_0 - 2 \ii \omega_0)^{-1} \hat D (\hat L_0 - 2 \ii \omega_0)^{-1} \bm n_2
             (\bm u,\bm u) ). \label{delta}
\end{align}
Moreover, calculation of $\langle z(\phi) \cdot \delta \bar w_0(\phi) {w'}_0^2(\phi')\rangle$ yields
the Fourier coefficients of $\Gamma$, given as
\begin{align}
 a_2 &= r^2 \re \delta (c_3-c_2), \label{a_2}\\
 b_2 &=-r^2 \re \delta ( 1+c_2 c_3), \label{b_2}
\end{align}
where $c_3=\im \delta /\re\delta$.
As discussed in Sec.~\ref{sec:cluster}, the stability of phase clusters
crucially depends on these Fourier coefficients.

\section{Demonstration}
Our theory is applied to the prediction of cluster states in globally
coupled oscillators. First, we
briefly summarize the existence and stability of cluster states in the
phase model with global coupling. We then numerically confirm our
prediction about clustering behavior in the Brusselator model.

\subsection{Existence and stability of balanced cluster states} \label{sec:cluster}
A globally coupled system with $N$ oscillators
is obtained by replacing $\bm x'$ in Eq.~(\ref{full_model2}) with
${\bm X} \equiv \frac{1}{N} \sum_{j=1}^N \bm x_j$. The corresponding phase models
given in Eq.~\eqref{pm}
with $\kappa$ being replaced by $\e^2 \kappa$.
By assuming $\kappa>0$ and rescaling time scale, we drop this $\kappa \e^2$.
For $\kappa<0$, all the eigenvalues given below
will have the opposite sign.

The balanced $n$-cluster state ($n\ge 2$) is the state in which
the whole population splits into equally populated $n$ groups
(here we assume that $N$ is a multiple of $n$),
oscillators in group $m$ $(m=0,1,\ldots,n)$ have an identical
phase $\psi_m$, and the phase of groups are equally separated
($\psi_m=\Omega t + 2m\pi/n$). 
In Eq.~(\ref{pm}), this solution always exists for any $n$.

The stability of the balanced cluster states was studied by Okuda
\cite{okuda93}. The $n$-cluster state with $n\ge 2$ possesses two
types of eigenvalues, which are associated with intra-cluster and
inter-cluster perturbations.  The eigenvalue associated with
intra-cluster perturbations is given by $\lambda^{\rm intra}_{n} =
\sum_{k=1}^\infty b_{k n}$. Therefore, in the absence of the
$\ell$-th harmonics with $\ell \ge n$ in $\Gamma$, the $n$-cluster state
has a zero eigenvalue. That is, the $n$-cluster state may not be asymptotically
stable.
This is the reason why any multi-phase clusters do not appear in the lowest order
amplitude equation given by Eq.~\eqref{gl}.
%
%

However, 
as discussed in Sec.~\ref{sec:higher_order}, the amplitude equation that
appropriately takes into account higher order corrections
yields higher harmonics in the corresponding phase coupling function
$\Gamma$. The
order of higher harmonics is given by Eq.~\eqref{order}, implying
$\lambda_{n}^{\rm intra}
=\sum_{k=1}^\infty b_{k n}=
O(\e^{2(n-1)})$.  Therefore, multi-phase clusters can be asymptotically stable.

%
%

From here, we focus on one- and two-cluster states.
By neglecting third and higher harmonics in $\Gamma$, eigenvalues for
the one-cluster ($\lambda_{1}^{\rm intra}$) and the balanced two-cluster
states $(\lambda_{2}^{\rm intra}, \lambda_{2}^{\rm inter})$ are given by
\begin{align}
 &\lambda_{1}^{\rm intra} = \Gamma'(0) = b_1+2 b_2, \label{l1}  \\
 &\lambda_{2}^{\rm intra} = \frac{1}{2} \left(
\Gamma'(0)+\Gamma'(\pi) \right)=2b_2,  \label{l2_intra} \\
&\lambda_{2}^{\rm inter} =
\Gamma'(\pi) = -b_1+2 b_2.  \label{l2_inter}  
\end{align}
By substituting 
Eqs.~\eqref{b_1} and \eqref{b_2}, we can obtain the expressions for these eigenvalues.
At this point, we should recall that these expressions for $b_1$ and
$b_2$ involve the errors of $O(\e^2)$ and
$O(\e^4)$, respectively. 
We should thus keep in mind the following estimation when we substitute the
expressions given by Eqs.~\eqref{b_1} and \eqref{b_2} into Eqs.~\eqref{l1}--\eqref{l2_inter}:
\begin{align}
 \lambda_{1}^{\rm intra} &=  b_1+ O(\e^2), \label{l1_order}  \\
 \lambda_{2}^{\rm intra} &= 2b_2 + O(\e^4),  \label{l2_intra_order} \\
\lambda_{2}^{\rm inter} &= -b_1 + O(\e^2).  \label{l2_inter_order}  
\end{align}
Equations \eqref{l1_order}--\eqref{l2_inter_order} imply that 
the parameter region in which the balanced two-cluster state is
asymptotically stable well coincide the region with 
$b_1>0$ and $b_2<0$.


There is another type of two-cluster states, in which the phase difference
between the clusters are different from $\pi$.
This type of two-cluster states is usually unstable.
However, a pair of two-cluster states may form attracting heteroclinic
cycles. In such a case, an interesting dynamical behavior called slow switching appears\cite{hansel93,
kori01, kori03}. Although 
the clusters are generally not equally populated in this type of two-cluster states,
we consider only two equally populated clusters for
simplicity in the present paper. 
For convenience, we refer to the two-cluster states with the phase
difference $\pi$ and other phase differences
as anti-phase and
out-of-phase cluster states, respectively.

For out-of-phase cluster states,
one may show that the phase difference $\Delta \phi$ between the clusters is given
by $\Delta \phi = \arccos (-b_1/2b_2)$. 
The solution to $\Delta \phi$ exists only
when $b_1$ is comparable to or even smaller than $b_2$.
As $b_1$ is generally much smaller than $b_2$,
this situation typically occurs
near the stability boundary of the one cluster state 
at which $b_1$ changes its sign.
%

There are three types of eigenvalues for out-of-phase cluster state, given as
\begin{align}
\lambda_{\rm ss}^{\rm (i)} & = \frac{1}{2} \left( \Gamma'(0)+\Gamma'(\Delta \phi)
 \right) = -\frac{a_1}{2} \sin \Delta \phi + O(\e^2), \label{lss_1}\\
\lambda_{\rm ss}^{\rm (ii)} &= \frac{1}{2} \left( \Gamma'(0)+\Gamma'(-\Delta \phi)
 \right) = \frac{a_1}{2} \sin \Delta \phi + O(\e^2),  \label{lss_2}\\
\lambda_{\rm ss}^{\rm (iii)} &= \frac{1}{2} \left( \Gamma'(\Delta \phi)+\Gamma'(-\Delta \phi)
 \right) = \frac{(b_1+2b_2)(b_1-2b_2)}{2b_2} = -\lambda_1 (1+\cos \Delta
 \phi). \label{lss_3}
\end{align}
Local stability conditions necessary for the
slow switching dynamics are \cite{hansel93,kori01}
$\lambda_{\rm ss}^{\rm (i)} \lambda_{\rm ss}^{\rm (ii)} < 0$, $\lambda^{\rm (iii)}_{\rm ss}<0$, and 
\begin{equation}
 \lambda_{\rm ss}^{\rm (i)}+\lambda_{\rm ss}^{\rm (ii)} = \frac{b_1(b_1+2b_2)}{2b_2} =
  \lambda_1 \cos \Delta \phi<0.
\end{equation}
As $a_1$ is generally of $O(1)$,
the first condition always holds true.
This means that any out-of-phase cluster state is saddle.
The second condition holds only when $\lambda_1 > 0$ (i.e., the one
cluster state is unstable). Then, the last condition is satisfied 
only when $\cos \Delta \phi < 0$, i.e., $b_1$ and $b_2$ have the same
sign. Then, $\lambda_1 =b_1 + 2 b_2 > 0$ implies $b_1>0$ and $b_2>0$.
Therefore, the slow switching dynamics may arise
only when $b_1>0$, $b_2>0$ and $b_1$ is comparable to $b_2$.


We summarize general properties of the cluster states. Here, we consider
both positive and negative coupling strength $\kappa$.
\begin{itemize}
 \item $\kappa b_1 <0$: one cluster state is stable, anti-phase cluster state
       is unstable.
\item $\kappa b_1 >0, \kappa b_2<0$: one cluster state is unstable, anti-phase cluster
      state is stable.
\item $\kappa b_1 >0, \kappa b_2>0$: one cluster state is unstable,
      balanced two cluster state is unstable,
slow switching may arise near the stability
boundary for one cluster state.
\end{itemize}


\subsection{Numerical verification with limit-cycle oscillators}
\begin{figure}
 \includegraphics[width=12cm]{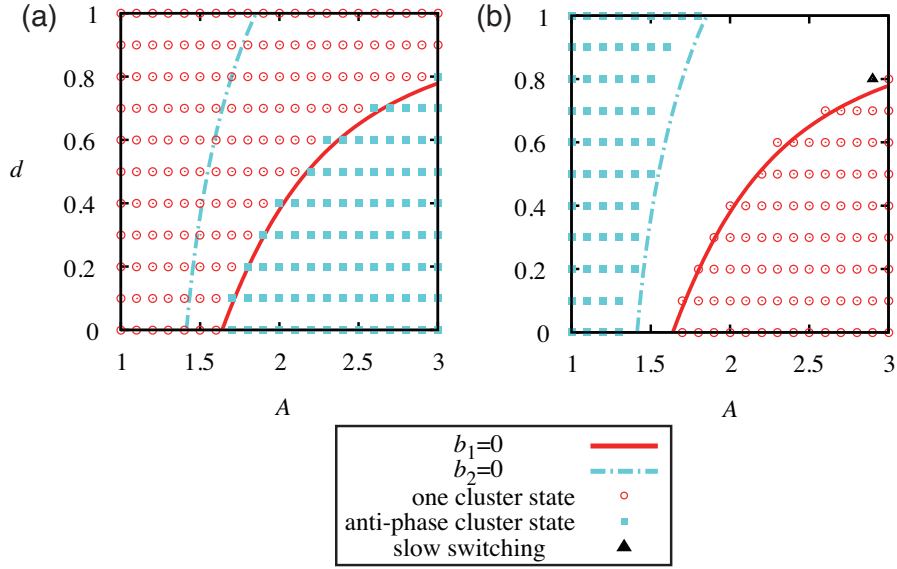}
 \caption{(color online) Phase diagram of cluster states in the
 Brusselator system for (a) positive ($\kappa = 0.001$) and (b) negative
 coupling ($\kappa = -0.001$). On solid and dashed lines, $b_1$ and $b_2$ change their signs,
 respectively. The regions $b_1>0$ and $b_2<0$ are right of the lines.
Numerical data are obtained by direct numerical simulations of Eqs. (\ref{bl1}) and
 (\ref{bl2}) with $N=4$,
 $B = (1+\e^2)B_{\rm c}, \e= r \sqrt{\frac{2+A^2}{A^2(1+A^2)}}, r=0.1$.}
\label{fig:phase_diagram}
\end{figure}

To verify our theory, we consider a population of Brusselator
oscillators with global coupling, 
whose dynamical equations are given by
\begin{subequations}
\label{bl}
\begin{align}
 \frac{dx_i}{dt} &= A-(B+1) x_i + x_i^2 y_i + \frac{\kappa}{N} \sum_{j=1}^N (x_j-x_i), \label{bl1} \\
 \frac{dy_i}{dt} &= B x_i - x_i^2 y_i + \frac{\kappa d}{N}
  \sum_{j=1}^N (y_j-y_i). \label{bl2}
\end{align}
\end{subequations}
We treat $B$ as a control parameter and fix other parameters $A$ and
$d$.
In the absence of coupling (i.e., $\kappa=0$),
the Hopf bifurcation occurs at $B=B_{\rm c} \equiv 1+A^2$. 
The bifurcation parameter is defined as
$\e^2 = \frac{B-B_{\rm c}}{B_{\rm c}}$ for $B>B_{\rm c}$. 
As shown in Appendix \ref{sec:appB},
using the expressions given in Sec.~\ref{sec:theory}, 
we obtained
Fourier coefficients $a_1, a_2, b_1$ and $b_2$ of $\Gamma$ as functions
of $A,B,d$ and $\e$. 
Figure \ref{fig:phase_diagram} displays the
phase diagram in the parameter space $(A,d)$, where the lines show the predicted
stability boundaries given by $b_1=0$ and $b_2=0$. 
Slow switching
dynamics is predicted to occur in the narrow region left of the line
$b_1=0$ for $\kappa<0$.

\begin{figure}
 \includegraphics[width=12cm]{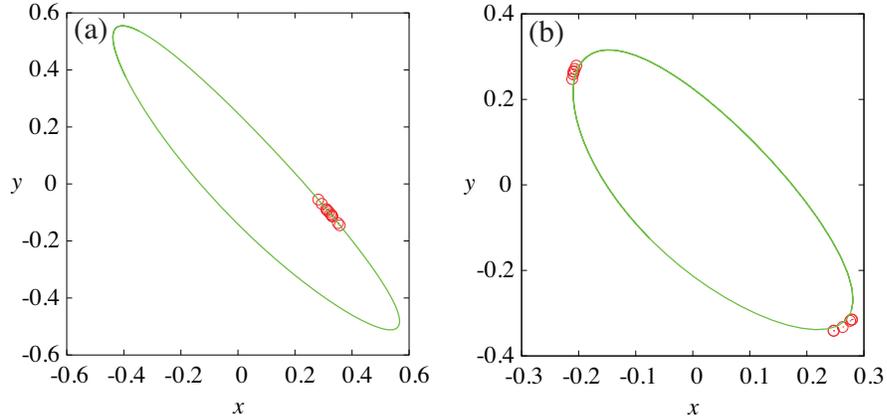}
 \caption{(color online) Snapshots of cluster states in the Brusselator system. For a
 better presentation, displayed snapshots are those before complete
 convergence. In (a), the one cluster state is obtained.
In (b), the balanced two-cluster state is obtained.
Parameter values: $N=12$, $\kappa=-0.001, d=0.4, B = (1+\e^2)B_{\rm c}, \e= r \sqrt{\frac{2+A^2}{A^2(1+A^2}}$, $r=0.1$, (a) $A=2.8$, (b) $A=1.1$.}
 \label{fig:snaps}
\end{figure}

We carried out direct numerical simulations of Eqs. (\ref{bl1}) and
(\ref{bl2}).
Starting from random initial conditions, the system typically converged
to balanced cluster states.
Two snapshots are displayed in Fig. \ref{fig:snaps}.
The symbols in Fig. \ref{fig:phase_diagram} display the parameter sets at
which the indicated cluster state is obtained.
We also found the slow switching dynamics at the filled triangle in
Fig. \ref{fig:phase_diagram}(b)], as theoretically expected.
We expect that the slow switching dynamics would appear anywhere
in the narrow region left to the line $b_1$ if the parameter $A$ and $d$
were more finely varied.
All together, we have an excellent agreement between the analytical and numerical results.

\begin{figure}
 \includegraphics[width=10cm]{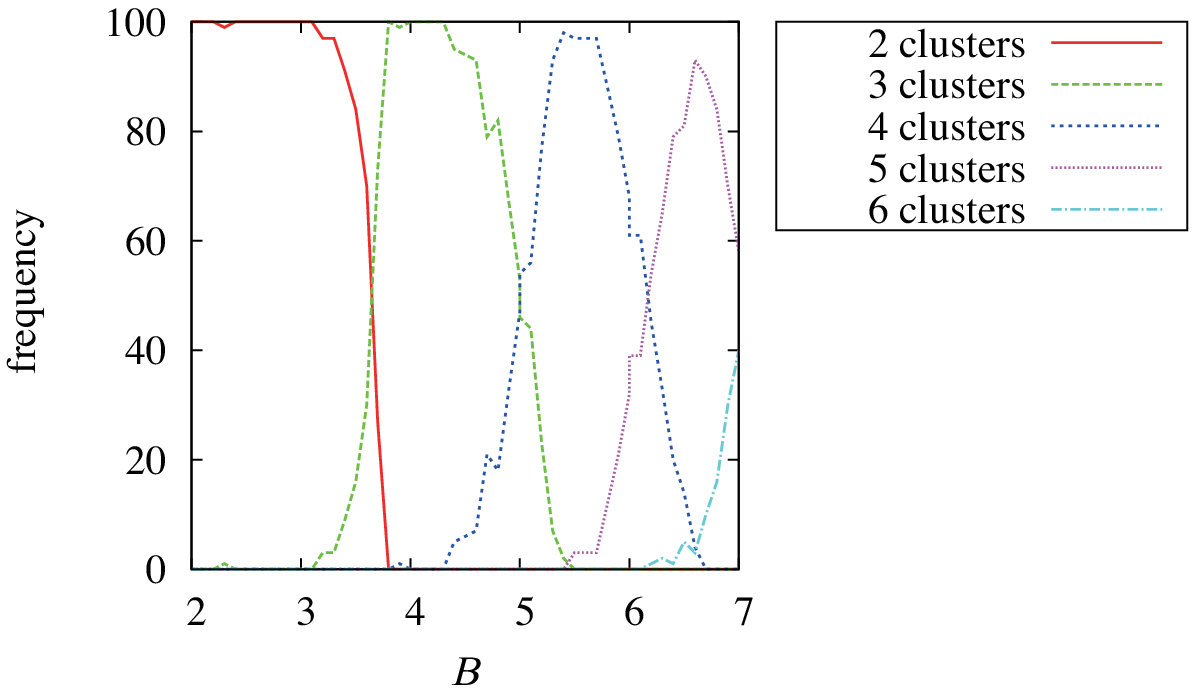}
 \caption{(color online) Clustering behavior in the Brusselator system
farther from the Hopf bifurcation point. Each line 
indicates how many times the $n$ cluster state (not necessarily
 perfectly balanced) was obtained out of 100 different initial
conditions. $N=24$, $A=1.0$, $d = 0.7$, $\kappa =-0.01$.
Initial values of $x_i$ and $y_i$ are random numbers
taken from the uniform distribution in the range $[-0.05,0.05]$.
}
 \label{fig:hist}
\end{figure}

We next observed clustering behavior for various $B$ values
far from the bifurcation point $B_{\rm c}$ with $N=24$ oscillators. 
We fixed $A=1.0$, so that $B_{\rm c}=2.0$. At each $B$ value, we employed 100 different random
initial conditions. For each initial condition, we checked the number of
phase clusters after transient time.  Figure \ref{fig:hist} shows 
the frequency of appearance of the $n$ cluster state for each $B$ value.
As the system is farther from the bifurcation point, $n$-cluster states
with larger $n$ values were more likely to appear. In this particular case, the
number of clusters tends to increase as the system is farther from the
bifurcation point. This result indicates that higher harmonics in phase coupling
function are developed as the bifurcation parameter increases, which is
consistent with our estimation given by Eq.~\eqref{order}.

\section{Experiments}

Experiments were conducted with a population of nearly identical $N=64$
electrochemical oscillators. Each oscillator is represented by a 1 mm diameter Ni wire embedded 
in epoxy and immersed in 3 mol/L sulfuric acid. The oscillators exhibit smooth or
relaxation waveforms for the current (the rate of dissolution),
depending on the applied potential (V) vs. a $\textrm{Hg/Hg}_2
\textrm{SO}_4/\textrm{cc.}\ \textrm{K}_2 \textrm{SO}_4$ reference electrode. 
The current of the electrodes
became oscillatory through a supercritical Hopf bifurcation point at
$V=1.0\textrm{V}$; the oscillations are smooth near the Hopf bifurcation
point. As the potential is increased, relaxation oscillations are seen
that disappear into a steady state through a homoclinic bifurcation at
about $V=1.31\textrm{V}$ \cite{zhai04}.  The parameters (applied
potential) were chosen such that the the oscillators exhibit smooth
oscillations near the Hopf bifurcation without any coupling. The
electrodes were then coupled with a combination of series ($R_{s}$) and
parallel ($R_{p}$) resistors such that the total resistance $ R_{tot}=R_p + 64 R_s$ is kept constant at 
$ 652 \textrm{Ohm}$. The imposed coupling strength can be
computed as $K=NR_{s}/R_{p}$ (More experimental details are given in
Ref.~\cite{zhai05}). Negative coupling was induced with the application
of negative series resistance supplied by a PAR 273A potentiostat in the form 
of IR compensation. 
The cluster states are obtained from nearly uniform initial conditions that correspond
to zero current (no metal dissolution).

\begin{figure}
\includegraphics[width=0.9\columnwidth]{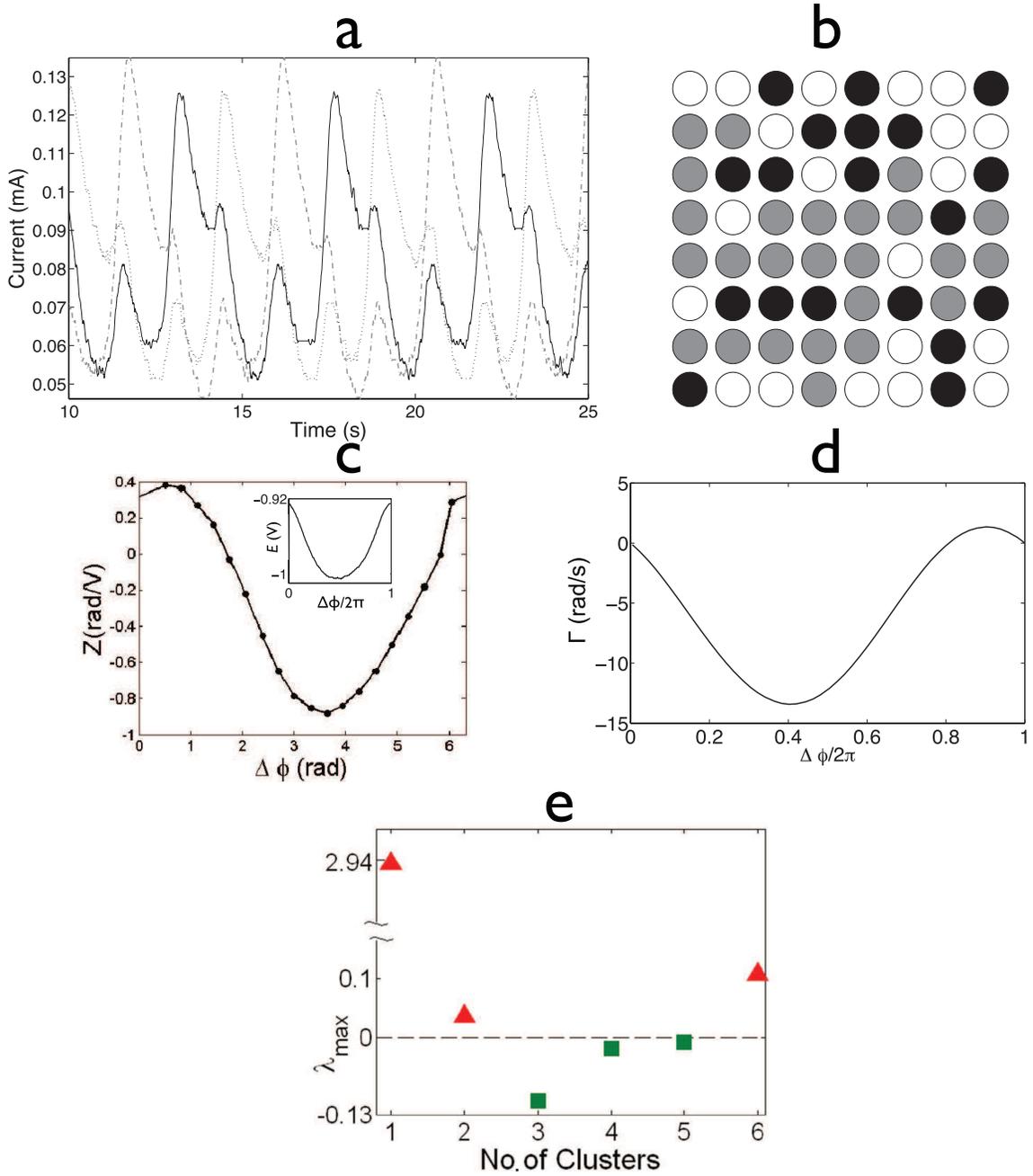}
\caption{(color online) Experiments: Three cluster state close to Hopf-bifurcation with negative
global coupling of 64 electrochemical oscillators. $ K=-0.88 $(a) Current
time series and the three cluster configuration at $V=1.05$ V (close
to a Hopf bifurcation.) Solid, dashed, and dotted curves represent currents from the three clusters.  b) Cluster configuration. (White, black, and gray circles represent the three clusters.) (c) Response function
and waveform (inset) of electrode potential $ E = V - I R_{tot}$, where $I$ is the current of the 
single oscillator. d) Phase coupling function. e) Stability analysis
of the clusters with experiment-based phase models for $K=-1$: the maximum of
the real parts of eigenvalues 
for each balanced cluster state.}
\label{fig:exp-three}
\end{figure}

Three cluster states were observed near the Hopf bifurcation (V = 1.05
V) with negative coupling. Fig. \ref{fig:exp-three}(a) shows the current
from one oscillator of each of the three clusters.
The nearly balanced three-cluster state with configuration
(25:20:19) is shown on a grid of 8x8 circles \ref{fig:exp-three}(b).
Each shade represents one cluster. 
In the previous work on the same system it had been shown that with
positive coupling close to the Hopf bifurcation only one-cluster state
is present \cite{kiss05}.  Phase response curves
(Fig. \ref{fig:exp-three}(c)) and coupling functions
(Fig. \ref{fig:exp-three}(d)) for these oscillators were determined
experimentally by introducing slight perturbations to the oscillations
\cite{kiss05}. The stability of these cluster states was determined by
computing the eigenvalues of the phase model \cite{okuda93}. The maxima
of real parts of these eigenvalues, for the same potential as in
Fig. \ref{fig:exp-three}(a), are shown in Fig. \ref{fig:exp-three}(e).
It is clear that with negative coupling multi-cluster states should be
observed and the three-cluster
state is the most stable state.

\begin{figure}
\includegraphics[width=0.9\columnwidth]{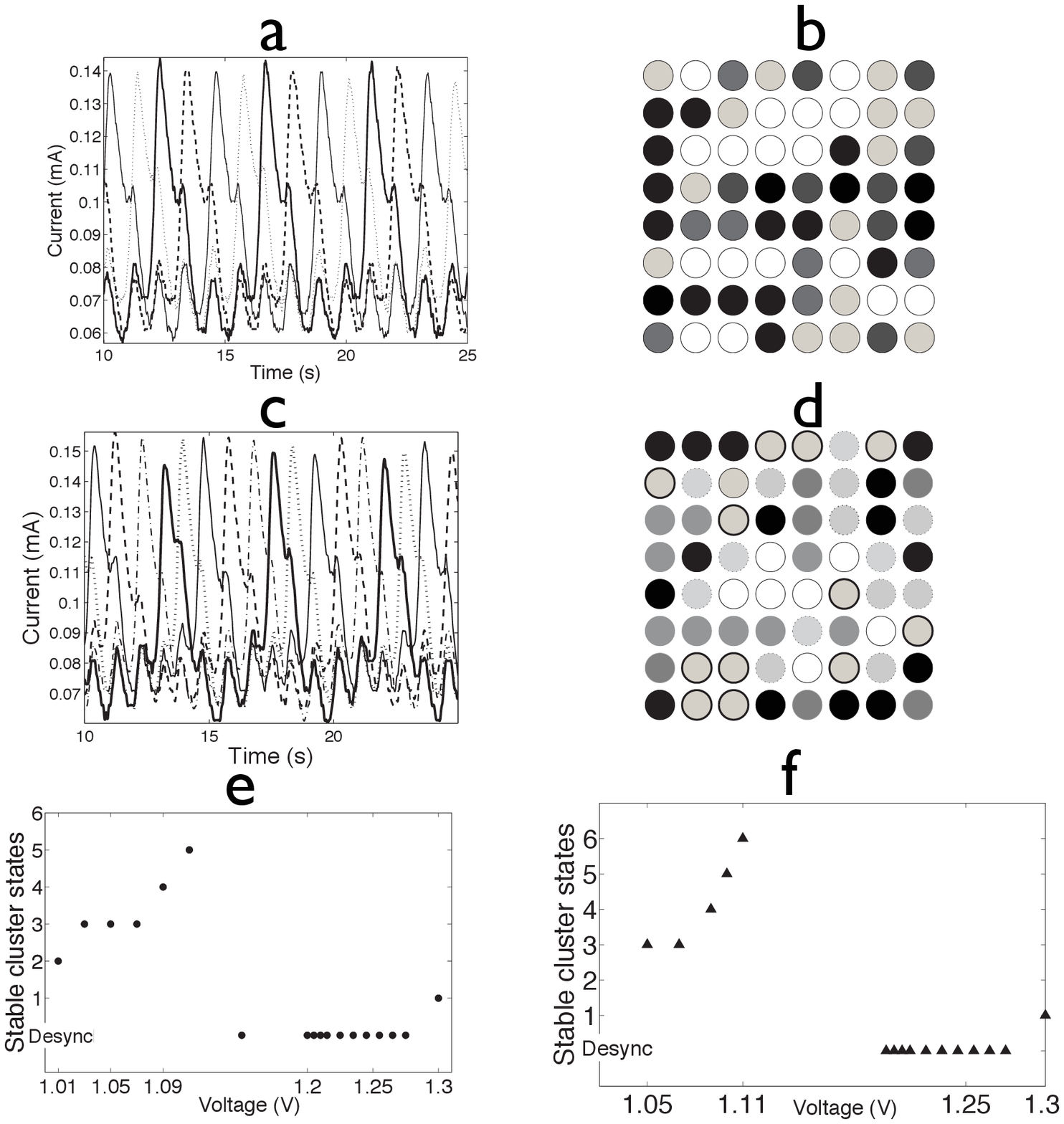}
\caption{Experiments: Multiple cluster states observed as the parameters are moved away from Hopf-bifurcation with
negative global coupling for 64 electrochemical oscillators.
Top row: four clusters, $V=1.09 V$, $K=-0.88$. Middle row: five clusters. $V=1.11 V$, $K=-0.88$
(a) Current time series of the four clusters.   (b) Cluster configuration
(17:14:15:18) for four cluster state. (c) Current time series.  d) Cluster configuration (15:13:14:7:15) of five cluster
state. e) Experimentally observed cluster states as a function of
applied potential. (f) The most stable cluster state as a function of potential
predicted by the experimentally obtained phase model. }
\label{fig:exp-many}
\end{figure}

As the potential was varied the number of cluster states changed. Four
and five cluster states were observed at higher potentials. Examples of
the oscillation waveforms and configurations for the 4 and 5 cluster
states are shown in Figs. \ref{fig:exp-many}(a), (b), (c), and
(d).
Further increase in the potential resulted in complete
desynchronization of the 64 oscillators.  At higher potentials, for
moderately relaxational oscillators, only one cluster state was
observed. Fig. \ref{fig:exp-many}(e) summarizes the effect of changing
the parameter (potential) on the existence of different cluster
states.
The presence of these clusters can be explained by the most
stable clusters from the experimentally determined phase model
(Fig. \ref{fig:exp-many}(f)). (We were not able to derive a phase model
for the two-cluster state because the amplitude of the oscillations was
too small for response functions to be measured in experiments.)

The experiments thus confirm that varying number of clusters (2-5) can
be observed in the electrochemical system close to Hopf bifurcation with
negative global coupling. Note that these clusters are different than
those reported previously that had been obtained with relaxation
oscillators with positive coupling \cite{zhai05}. Similar to the results
obtained with the Brusselator model, when the system is shifted farther
away from the Hopf bifurcation, the number of clustered increased due to
the emergence of stronger higher harmonics in the coupling function. In
agreement with the theory, the clusters required relatively strong
negative coupling ($K \approx -0.88 $) in contrast with the one cluster
state with positive coupling that required very weak coupling ($K < 0.05
$) \cite{kiss02}. Therefore, we see that weak higher harmonics can play
important role in determining the dynamical features of cluster
formation when the contribution of dominant harmonics does not induce a
stable structure.

\section{Concluding Remarks}
In summary, we have shown theoretically and confirmed numerically and experimentally
the development of higher harmonics in the phase coupling function and
the appearance of phase clusters in globally
coupled oscillatory systems.
We found that the only relevant higher-order terms that should be
included in the amplitude equation for weakly coupled oscillators
are $\bar w^{\ell-1} w'^{\ell}$ with
$\ell\ge 2 $. In particular, we derived the expression for the
coefficient of the term $\bar w w'^2$, which has, to our surprise, a
very concise form.  The relevance of higher harmonics in the phase
coupling function has been well recognized.  Our study uncovered how
higher harmonics are developed in limit-cycle oscillators near a Hopf
bifurcation point.  The derived amplitude equation
will serve as an analytically tractable limit-cycle oscillator model
that produces higher harmonics in the phase coupling function.

\clearpage
\appendix

\section{Derivation of the amplitude equation} \label{sec:appA}
Our aim is to reduce Eq.~\eqref{full_model2} to
the amplitude equation given by
Eq.~\eqref{gl2}. Because the expressions for $\alpha, \beta$ and
$\gamma$ are obtained in Ref.~\onlinecite{kuramoto84}, we focus on $\delta$.

For convenience, we rewrite Eq.~\eqref{full_model2} and 
Eq.~\eqref{reduction_w} as
\begin{equation}
\dot{\bm x} = \hat L_0 \bm{x}+\e^2 \hat L_1
  \bm{x}+\bm{n}_2(\bm{x},\bm{x})+\bm{n}_3(\bm{x},\bm{x},\bm{x})+
  \epsilon^2 \kappa \hat D \bm x',
  \label{full_model3}
\end{equation}
\begin{equation}
 \dot w = G(w,\bar w, w', \bar w'),
 \label{dot_w}
\end{equation}
respectively.
Here, $\bm n_2$ and $\bm n_3$ are defined as
\begin{align}
 \bm n_2(\bm u, \bm v) &= \sum_{i,j=1}^M \frac{1}{2!}
  \left(\frac{\partial^2 \bm f }{\partial x_i \partial x_j} \right)_
  {\bm x= \bm 0} u_i v_j,\\
 \bm n_3(\bm u, \bm v, \bm w) &= \sum_{i, j,k=1}^M \frac{1}{3!}
  \left(\frac{\partial^3 \bm f }{\partial x_i \partial x_j \partial x_k} \right)_
  {\bm x= \bm 0} u_i v_j w_k,
\end{align}
where $\bm x=(x_1, x_2, \ldots, x_M)^{\rm T}$ and similar definitions
are applied to $\bm u, \bm v$ and $\bm w$.
Note that we consider only linear coupling in Eq.~\eqref{full_model3}. 
In the presence of nonlinear coupling,
the expression for $\delta$ will be different from Eq.~\eqref{delta}
while $\alpha, \beta$ and $\gamma$ are unchanged.

By substituting Eq.~\eqref{reduction_x} into Eq.~(\ref{full_model3}) and
using Eq.~\eqref{dot_w}, we obtain
\begin{equation}
{\mathscr L}_0\bm{\rho}=G\exp(\ii\theta)\bm{u}+\bar{G}\exp(-\ii\theta)\bar{\bm u}
+\bm{b}(w,\bar{w},w',\bar{w}',\theta),
\label{inhomogeneous}
\end{equation}
where
\begin{eqnarray}
{\mathscr L}_0&=&\hat L_0-\omega_0\frac{\partial}{\partial\theta},\\
\bm{b}&=&-\epsilon^2 \hat L_1\bm{x}-\bm{n}_2(\bm{x},\bm{x})-\bm{n}_3(\bm{x},\bm{x},
\bm{x})-\epsilon^2 \kappa \hat D\bm{x}' \nonumber \\
&+&G\frac{\partial\bm{\rho}}{\partial w}+\bar{G}\frac{\partial\bm{\rho}}{\partial\bar{w}}+
G'\frac{\partial\bm{\rho}}{\partial w'}+\bar{G}'\frac{\partial\bm{\rho}}
{\partial\bar{w}'}. \label{b}
\end{eqnarray}
Regard Eq.~(\ref{inhomogeneous}) formally as an inhomogeneous linear differential equation
for $\bm{\rho}(\theta)$, 
where the right-hand side
as a whole represents the inhomogeneous term. 
To solve Eq.~(\ref{inhomogeneous}),
$\bm{\rho}(\theta)$ and $\bm{b}(\theta)$ are expanded as
\begin{eqnarray}
\bm{\rho}(\theta)&=&\sum_{\ell=-\infty}^{\infty}\bm{\rho}^{(\ell)}\exp({\rm
 i}\ell\theta), \label{rho_l} \\
\bm{b}(\theta)&=&\sum_{\ell=-\infty}^{\infty}\bm{b}^{(\ell)}\exp({\rm
i}\ell\theta). \label{b_l}
\end{eqnarray}
Note that the terms $\exp(i\theta)\bm{u}$ and its complex conjugate in
Eq.~\eqref{inhomogeneous} 
are the zero-eigenvectors of ${\mathscr L}_0$; i.e., ${\mathscr
L}_0 (e^{\ii \theta} \bm u) = {\mathscr
L}_0 (e^{-\ii \theta} \bar {\bm u}) = 0$.
Because the left-hand side
in Eq.~(\ref{inhomogeneous}) is free of the zero-eigenvector components
due to the operation of ${\mathscr L}_0$,
these components must be canceled 
in the right-hand side as well. This condition is called the solvability condition.
By substituting Eqs.~\eqref{rho_l}
and \eqref{b_l}
into Eq.~\eqref{inhomogeneous}, and comparing the component of $\exp(\ii
\theta)$ in both sides, we obtain the solvability condition 
\begin{equation}
 G=-\bm{v}\bm{b}^{(1)}.
  \label{solvability}
\end{equation}
Further, by comparing other components, we obtain
\begin{eqnarray}
\bm{\rho}^{(\ell)}&=&(\hat L_0-{\rm
 i}\ell\omega_0)^{-1}\bm{b}^{(\ell)},\quad(\ell \ne \pm 1),\\
\bm{\rho}^{(1)}
&=&(\hat L_0-{\rm i}\omega_0)^{-1}(\bm{b}^{(1)} + G \bm u),\\
\bm{\rho}^{(-1)}&=&(\hat L_0+{\rm i}\omega_0)^{-1}(\bm{b}^{(-1)} + \bar G \bar {\bm u}).
\end{eqnarray}
Let $\bm{b}^{(\ell)}$ and $\bm{\rho}^{(\ell)}$ be further expanded in
the powers of
$\epsilon$:
\begin{eqnarray}
\bm{b}^{(\ell)} &=&
 \sum_{\nu=2}^{\infty} \epsilon^{\nu} \bm{\tilde b}_{\nu}^{(\ell)} 
= \sum_{\nu=2}^{\infty} \bm{b}_{\nu}^{(\ell)} ,\\
\bm{\rho}^{(\ell)}&=&
 \sum_{\nu=2}^{\infty}\epsilon^{\nu}\bm{\tilde \rho}_{\nu}^{(\ell)}
=  \sum_{\nu=2}^{\infty}\bm{\rho}_{\nu}^{(\ell)}.
\end{eqnarray} 
Correspondingly, $\bm{b}$ and $\bm{\rho}$ themselves are expanded as
\begin{eqnarray}
\bm{b} &=& \sum_{\nu=2}^{\infty}\epsilon^{\nu}\bm{\tilde b}_{\nu}
= \sum_{\nu=2}^{\infty}\bm{b}_{\nu},\\
\bm{\rho} &=& \sum_{\nu=2}^{\infty}\epsilon^{\nu}\bm{\tilde \rho}_{\nu}
= \sum_{\nu=2}^{\infty}\bm{\rho}_{\nu}.
\end{eqnarray}
Let $G$ be also expanded as
\begin{equation}
G=\sum_{\nu=1}^{\infty}\epsilon^{2\nu+1}\tilde G_{2\nu+1}
 = \sum_{\nu=1}^{\infty}G_{2\nu+1}.
\end{equation}
where we have anticipated the absence of even powers.

To derive Eq.~\eqref{gl2}, we need to calculate $G_3$ and $G_5$.
As $G_3$ is already obtained in Ref.~\onlinecite{kuramoto84}, 
our main concern is $G_5$, especially the 
higher order coupling term in $G_5$.
To obtain $G_3$ and $G_5$, we need the expressions for $\bm{b}_\nu$ $(\nu=1,\ldots,5)$. 
Because $\bm x_0 = O(\e)$; $\bm b_\nu, \bm \rho_\nu = O(\e^\nu)$ ($\nu \ge
2$); $G_\nu = O(\e^\nu)$ ($\nu \ge 3$), we find
\begin{eqnarray}
\bm{b}_2&=&-\bm{n}_2(\bm{x}_0,\bm{x}_0),\\
\bm{b}_3&=&-\e^2 \hat L_1\bm{x}_0-2\bm{n}_2(\bm{x}_0,\bm{\rho}_2)
 -\bm{n}_3(\bm{x}_0,\bm{x}_0,\bm{x}_0) -\kappa \e^2 \hat D\bm{x}_0',\\
\bm{b}_4&=&-\e^2 \hat L_1\bm{\rho}_2-2\bm{n}_2(\bm{x}_0,\bm{\rho}_3)
-\bm{n}_2(\bm{\rho}_2,\bm{\rho}_2)\nonumber\\
&&-3\bm{n}_3(\bm{x}_0,\bm{x}_0,\bm{\rho}_2)
-\kappa \e^2 \hat D\bm{\rho}_2',\\
\bm{b}_5&=&-\e^2 \hat L_1\bm{\rho}_3-2\bm{n}_2(\bm{x}_0,\bm{\rho}_4)
-2\bm{n}_2(\bm{\rho}_2,\bm{\rho}_3)\nonumber\\
&&-3\bm{n}_3(\bm{x}_0,\bm{x}_0,\bm{\rho}_3)
-3\bm{n}_3(\bm{x}_0,\bm{\rho}_2,\bm{\rho}_2) -\kappa \e^2 \hat D\bm{\rho}_3'\nonumber\\
&&+G_3\frac{\partial\bm{\rho}_2}{\partial w}+\bar{G}_3\frac{\partial 
\bm{\rho}_2}{\partial\bar{w}}+G_3'\frac{\partial\bm{\rho}_2}{\partial w'}
+\bar{G}_3'\frac{\partial\bm{\rho}_2}{\partial\bar{w}'}.
\end{eqnarray}

We first calculate $G_3 = -\bm v \bm{b}_3^{(1)}$. Using
\begin{eqnarray}
\bm{b}_3^{(1)} &=& -\e^2 \hat L_1\bm{x}_0^{(1)}-2\bm{n}_2(\bm{x}_0,\bm{\rho}_2)^{(1)}
 -\bm{n}_3(\bm{x}_0,\bm{x}_0,\bm{x}_0)^{(1)} -\kappa \e^2 \hat
 D\bm{x}_0'^{(1)} \nonumber \\
 &=& -\e^2 \hat L_1\bm{x}_0^{(1)}
 - 2\bm{n}_2(\bm{x}_0^{(1)},\bm{\rho}_2^{(0)})
  - 2\bm{n}_2(\bm{x}_0^{(-1)},\bm{\rho}_2^{(2)}) \nonumber \\
 && -3 \bm{n}_3(\bm{x}_0^{(1)},\bm{x}_0^{(1)},\bm{x}_0^{(-1)})
 -\kappa \e^2 \hat D\bm{x}_0'^{(1)}, \\
\bm{\rho}_2^{(0)} &=&  \hat L_0^{-1} \bm b_2^{(0)} \nonumber \\
 &=& - 2 \hat L_0^{-1} \bm n_2(\bm x_0^{(1)},\bm x_0^{(-1)}),  \\
\bm{\rho}_2^{(2)} &=&  (\hat L_0-2 \ii \omega_0)^{-1} \bm b_2^{(2)}
\nonumber \\
 &=& (\hat L_0-2 \ii \omega_0)^{-1} \bm n_2(\bm x_0^{(1)},\bm x_0^{(1)}),
\end{eqnarray}
we obtain Eq.~\eqref{gl} with Eqs.~\eqref{alpha}--\eqref{gamma}.

Now we calculate $G_5 = -\bm v \bm{b}_5^{(1)}$. We have 
\begin{align}
  \bm{b}_5^{(1)}
  = 
  &- \epsilon^2 \hat{L}_1 \bm{\rho}_3^{(1)}
  \nonumber \\
  &- 2 \bm{n}_2( \bm{x}_0, \bm{\rho}_4 )^{(1)}
  - 2 \bm{n}_2( \bm{\rho}_2, \bm{\rho}_3 )^{(1)}
  \nonumber \\
  &- 3 \bm{n}_3( \bm{x}_0, \bm{x}_0, \bm{\rho}_3 )^{(1)}
  - 3 \bm{n}_3( \bm{x}_0, \bm{\rho}_2, \bm{\rho}_2 )^{(1)}
  \nonumber \\
  &-\kappa \epsilon^2 \hat{D} \bm{\rho}_3'^{(1)}
  \nonumber \\
  &+ G_3 \frac{\partial \bm{\rho}_2^{(1)}}{\partial w}
  + \bar{G}_3 \frac{\partial \bm{\rho}_2^{(1)}}{\partial \bar{w}}
  + G_3' \frac{\partial \bm{\rho}_2^{(1)}}{\partial w'}
  + \bar{G}_3' \frac{\partial \bm{\rho}_2^{(1)}}{\partial \bar{w}'}
  \nonumber \\
  = 
  &- \epsilon^2 \hat{L}_1 \bm{\rho}_3^{(1)}
  \nonumber \\
  &- 2 \bm{n}_2( \bm{x}_0^{(1)}, \bm{\rho}_4^{(0)} )
  - 2 \bm{n}_2( \bm{x}_0^{(-1)}, \bm{\rho}_4^{(2)} )
  \nonumber \\
  &- 2 \bm{n}_2( \bm{\rho}_2^{(2)}, \bm{\rho}_3^{(-1)} )
  - 2 \bm{n}_2( \bm{\rho}_2^{(1)}, \bm{\rho}_3^{(0)} )
  - 2 \bm{n}_2( \bm{\rho}_2^{(0)}, \bm{\rho}_3^{(1)} )
  - 2 \bm{n}_2( \bm{\rho}_2^{(-1)}, \bm{\rho}_3^{(2)} )
  - 2 \bm{n}_2( \bm{\rho}_2^{(-2)}, \bm{\rho}_3^{(3)} )
  \nonumber \\
  &- 3 \bm{n}_3( \bm{x}_0^{(1)}, \bm{x}_0^{(1)}, \bm{\rho}_3^{(-1)} )
  - 6 \bm{n}_3( \bm{x}_0^{(1)}, \bm{x}_0^{(-1)}, \bm{\rho}_3^{(1)} )
  - 3 \bm{n}_3( \bm{x}_0^{(-1)}, \bm{x}_0^{(-1)}, \bm{\rho}_3^{(3)} )
  \nonumber \\
  &- 6 \bm{n}_3( \bm{x}_0^{(1)}, \bm{\rho}_2^{(2)}, \bm{\rho}_2^{(-2)} )
  - 6 \bm{n}_3( \bm{x}_0^{(1)}, \bm{\rho}_2^{(1)}, \bm{\rho}_2^{(-1)} )
  - 3 \bm{n}_3( \bm{x}_0^{(1)}, \bm{\rho}_2^{(0)}, \bm{\rho}_2^{(0)} )
  \nonumber \\
  &- 6 \bm{n}_3( \bm{x}_0^{(-1)}, \bm{\rho}_2^{(2)}, \bm{\rho}_2^{(0)} )
  - 3 \bm{n}_3( \bm{x}_0^{(-1)}, \bm{\rho}_2^{(1)}, \bm{\rho}_2^{(1)} )
  \nonumber \\
  &-\kappa \epsilon^2 \hat{D} \bm{\rho}_3'^{(1)}
  \nonumber \\
  &+ G_3 \frac{\partial \bm{\rho}_2^{(1)}}{\partial w}
  + \bar{G}_3 \frac{\partial \bm{\rho}_2^{(1)}}{\partial \bar{w}}
  + G_3' \frac{\partial \bm{\rho}_2^{(1)}}{\partial w'}
  + \bar{G}_3' \frac{\partial \bm{\rho}_2^{(1)}}{\partial \bar{w}'}.
\end{align}
Out of the above terms, we select those which produce
$\kappa \e^2 \hat Dw'^{2}\bar{w}$. Checking term by term, we find that
the following terms may safely be excluded:
\begin{itemize}
\item  those which include $\bm{\rho}_2$
\item  those which include $\bm{x}_0^{(1)}$
\item  those which include $\bm{x}_0^{(-1)}$ twice.
\end{itemize}
The remaining terms are
\begin{equation}
-\e^2 \hat L_1\bm{\rho}_3^{(1)}-2\bm{n}_2(\bm{x}_0^{(-1)},\bm{\rho}_4^{(2)})
-\kappa \e^2 \hat D\bm{\rho}_3'^{(1)}.
\end{equation}
The first of the above three terms is further dropped
because the coupling term included there is linear. 
The last term is also dropped because
the cubic term $\bm n_3(\bm x_0', \bm x_0', \bm x_0')$ yields
neither $w$ nor $\bar w$.
Thus, the only relevant term in $\bm b_5^{(1)}$ is the $\kappa$-dependent term in
\begin{equation}
-2\bm{n}_2(\bm{x}_0^{(-1)},\bm{\rho}_4^{(2)}).
 \label{eq0}
\end{equation}
The $\kappa$-dependent term in $\bm{\rho}_4^{(2)}$ is
\begin{equation}
(\hat L_0-2\ii\omega_0)^{-1}(-\kappa \e^2 \hat D\bm{\rho}_2'^{(2)}).
\label{eq1}
\end{equation}
Because
\begin{eqnarray}
\bm{\rho}_2'^{(2)}&=&(\hat L_0-2\ii\omega_0)^{-1}\big(-\bm{n}_2(\bm{x}_0'^{(1)},
\bm{x}_0'^{(1)})\big)\nonumber\\
&=&-w'^2(\hat L_0-2\ii\omega_0)^{-1}\bm{n}_2(\bm{u},\bm{u}),
\end{eqnarray}
Eq.~\eqref{eq1} becomes
\begin{equation}
\kappa \e^2 w'^2(\hat L_0-2\ii\omega_0)^{-1}\hat D(\hat L_0-2\ii\omega_0)^{-1}\bm{n}_2(\bm{u},\bm{u}).
\end{equation}
Thus, the relevant term in Eq.~\eqref{eq0} is
\begin{equation}
-2\kappa \e^2 w'^2\bar{w}\bm{n}_2\big(\bar{\bm{u}},(\hat L_0-2\ii\omega_0)^{-1}\hat D(\hat L_0-2\ii\omega_0)^{-1}
\bm{n}_2(\bm{u},\bm{u}) \big),
\end{equation}
which yields $\delta$ shown in Eq.~\eqref{delta}.

\section{Amplitude equation for the Brusselator model} \label{sec:appB}
We derive the expression for $\alpha, \beta, \gamma$ and $\delta$ for the
Brusselator model given by Eq.\eqref{bl}.
There are three parameters, $A$, $B$ and $d$, in Eq.\eqref{bl}.
We consider $B$ as a bifurcation parameter while $A$ and $d$ are fixed,
so that the expression for $\alpha,\beta, \gamma$ and $\delta$ will be
functions of $A$ and $d$. Note that 
such expressions except for 
$\delta$ were already derived in
Ref.~\onlinecite{kuramoto84}. 

The steady solution to Eq.~\eqref{bl} is $(x_0, y_0)=(a, b/a)$.
Introducing $\xi=x-x_0$ and $\eta=y-y_0$ and 
substituting them into Eq.~\eqref{bl}, we obtain
\begin{subequations}
\begin{align}
 \frac{d\xi_i}{dt} &= (B-1) \xi_i + A^2 \eta_i + f(\xi_i,\eta_i) + \frac{\kappa}{N} \sum_{j=1}^N (\xi_j-\xi_i), \label{bl3} \\
 \frac{d\eta_i}{dt} &= -B \xi_i - A^2 \eta_i - f(\xi_i,\eta_i)  + \frac{\kappa d}{N}
  \sum_{j=1}^N (\eta_j-\eta_i), \label{bl4}
\end{align}
\label{bl_change}
\end{subequations}
where
\begin{equation}
 f(\xi,\eta) = \frac{B}{A} \xi^2 + 2 A \xi \eta + \xi^2 \eta.
\end{equation}
In the absence of coupling (i.e., $\kappa=0$),
the trivial solution $(\xi_i, \eta_i)=(0,0)$ undergoes a supercritical
Hopf bifurcation at $B=B_{\rm c} \equiv 1+A^2$. We define the bifurcation
parameter as $\e^2 = \frac{B-B_{\rm c}}{B_{\rm c}}$.
We then obtain
\begin{equation}
 \hat L_0 = \left(
    \begin{array}{cc}
      A^2 & A^2 \\
      -(1+A^2) & -A^2
    \end{array}
  \right),
\end{equation}
\begin{equation}
 \hat L_1 = (1+A^2) \left(
    \begin{array}{cc}
      1 & 0 \\
      -1 & 0
    \end{array}
  \right),
\end{equation}
\begin{equation}
 \hat D = \left(
    \begin{array}{cc}
      1 & 0 \\
      0 & d
    \end{array}
  \right),
\end{equation}
\begin{equation}
 \bm u = \left(
    \begin{array}{c}
      1  \\
      -1 + \ii A^{-1}
    \end{array}
  \right),
\end{equation}
\begin{equation}
 \bm v = \frac{1}{2} \left(1-\ii A \;\; -\ii A \right),
\end{equation}
\begin{equation}
 \omega_0  = A,
\end{equation}
\begin{equation}
 \hat L_0^{-1} = \frac{1}{A^2} \left(
    \begin{array}{cc}
      -A^2  & -A^2 \\
     A^2+1 & A^2
    \end{array}
  \right),
\end{equation}
\begin{equation}
 (\hat L_0-2\ii\omega_0)^{-1} = \frac{1}{3A^2} \left(
    \begin{array}{cc}
      A^2 + 2 \ii A  & 3 A^2 \\
      -A^2 - 1 & -A^2 - 2 \ii A
    \end{array}
  \right).
\end{equation}
We introduce $\bm u_i = (\sigma_i, \mu_i)^{\rm T}$ ($i=1,2,3$) and write
\begin{equation}
 \bm n_2 ( \bm u_1, \bm u_2) =\left\{
  \frac{1+A^2}A \sigma_1 \sigma_2 + A(\sigma_1 \mu_2 + \mu_1 \sigma_1)
\right\}
  \left(
    \begin{array}{c}
      1 \\
      -1
    \end{array}
  \right),
\end{equation}
\begin{equation}
 \bm n_3 ( \bm u_1, \bm u_2, \bm u_3) =
  \frac{\sigma_1 \sigma_2 \mu_3 + \mu_1 \sigma_2 \sigma_3 +
  \sigma_1 \mu_2 \sigma_3}{3}
  \left(
    \begin{array}{c}
      1 \\
      -1
    \end{array}
  \right).
\end{equation}

Substituting these expressions to Eqs.~\eqref{alpha}--\eqref{gamma} and \eqref{delta}, we obtain
\begin{eqnarray}
 \alpha &=& \frac{1}{2} + \frac{A^2}{2}, \\
 \beta &=& \frac{1}{A^2} + \frac{1}{2} + \frac{\ii}{2}\left( \frac{4}{3A^3} -
					     \frac{7}{3A} + \frac{4A}{3}
					    \right),\\
\gamma &=& \frac{1}{2} + \frac{d}{2} + \frac{\ii}{2} \left(-A + Ad \right),\\
 \delta &=& -\frac{8}{3}+\frac{4}{9 A^6}+\frac{8}{3
  A^4}-\frac{1}{A^3}+\frac{28}{9 A^2}+\frac{1}{A}+2 A-\frac{32
  d}{3}+\frac{4 d}{9 A^6}-\frac{68 d}{9 A^2} + \nonumber \\
&&\ii \left(2+\frac{4}{9 A^5}+\frac{4}{A^3}+\frac{2}{A^2}+\frac{88}{9 A}+\frac{16 A}{3}+\frac{14 d}{9 A^5}+\frac{6 d}{A^3}+\frac{20 d}{9 A}-\frac{16 A d}{3}\right).
\end{eqnarray}
We further obtain
\begin{align}
c_1 &= \frac{\gamma_{\rm I}}{\gamma_{\rm R}}= -\frac{A(1-d)}{1+d},\\
c_2 &= \frac{\beta_{\rm I}}{\beta_{\rm R}}=
\frac{4-7A^2 + 4A^4}{3A(2+A^2)},\\
c_3 &= \frac{\delta_{\rm I}}{\delta_{\rm R}}=
\frac{A(4+5 d + (2-11d)A^2 + (d-1)A^4 )}  {4+d +
(2-10d)A^2 + (7d-2)A^4},\\
r &= \e \sqrt{\frac{\alpha_{\rm R}}{\beta_{\rm R}}}= \e \sqrt{\frac{A^2(1+A^2)}{2+A^2}}.
\end{align}
Using these coefficients, it is straightforward to obtain the expression
for the Fourier coefficients $a_1,a_2,b_1$ and $b_2$ of $\Gamma$.

\end{document}